\documentclass[journal]{IEEEtran}
\usepackage{cite}

\ifCLASSINFOpdf
   \usepackage[pdftex]{graphicx}
   \DeclareGraphicsExtensions{.pdf,.jpeg,.png}
\else
   \usepackage[dvips]{graphicx}
   \DeclareGraphicsExtensions{.eps}
\fi
\usepackage{array}
\usepackage{url}

\usepackage[utf8]{inputenc}

\usepackage{cleveref}
\usepackage{amsmath}

\usepackage{color}
\usepackage{eurosym}
\usepackage{amsfonts}
\usepackage{amsthm}
\usepackage{amssymb}
\usepackage{mathtools}
\usepackage{tikz}
\usepackage{textcomp}
\usepackage{multirow}
\usepackage{stmaryrd}
\usepackage[T1]{fontenc}
\usepackage{csquotes}
\usetikzlibrary{arrows}

\usetikzlibrary{matrix,calc,shapes}
\tikzset{
  treenode/.style = {shape=rectangle,
                     draw, anchor=center,
                     text width=9em, align=center,
                     top color=white, bottom color=white,
                     inner sep=1ex},
  decision/.style = {treenode, diamond, rounded corners, inner sep=0pt},
  root/.style     = {treenode, font=\Large, bottom color=red!30},
  env/.style      = {treenode, font=\ttfamily\normalsize},
  finish/.style   = {root, bottom color=green!40},
  dummy/.style    = {circle,draw}
}
\newcommand{\yes}{edge node [left] {yes}}

\DeclarePairedDelimiterX{\norm}[1]{\lVert}{\rVert}{#1}

\newcommand{\sumP}{\ensuremath{\displaystyle\sum_{m\in\omega_n} P_{nm}}}

\newcommand{\cnm}{\ensuremath{\displaystyle\sum_{g\in\mathcal{G}} c_n^g\gamma_{nm}^g}}
\newcommand{\TradingCost}{\ensuremath{\displaystyle\sum_{m\in\omega_n} c_{nm}P_{nm}}}

\def\be{\begin{equation}}
\def\ee{\end{equation}}
\def\barr{\begin{array}}
\def\earr{\end{array}}

\begin{document}
%
\title{Consensus-based Approach to Peer-to-Peer Electricity Markets with Product Differentiation}
%
%
%

\author{Etienne Sorin, Lucien Bobo, Pierre Pinson, \IEEEmembership{Senior Member,~IEEE}%

\thanks{Manuscript received ??, 2017; revised ??, 2017.}
\thanks{The authors are with the Technical University of Denmark, Department of Electrical Engineering, Kgs.\ Lyngby, Denmark (emails: \{egsorin,ppin\}@elektro.dtu.dk).}%
\thanks{The authors are partly supported by the Danish Innovation Fund and the ForskEL programme through the projects '5s' - Future Electricity Markets (12-132636/DSF), CITIES (DSF-1305-00027B) and The Energy Collective (grant no. 2016-1-12530).}
}

\maketitle

%
%

\markboth{IEEE Transactions On Power Systems,~Vol.~x, No.~x}{Sorin \MakeLowercase{\textit{et al.}}: Consensus-based Approach to Peer-to-Peer Electricity Markets }
%




\begin{abstract}
With the sustained deployment of distributed generation capacities and the more proactive role of consumers, power systems and their operation are drifting away from a conventional top-down hierarchical structure. Electricity market structures, however, have not yet embraced that evolution. Respecting the high-dimensional, distributed and dynamic nature of modern power systems would translate to designing peer-to-peer markets or, at least, to using such an underlying decentralized structure to enable a bottom-up approach to future electricity markets. A peer-to-peer market structure based on a Multi-Bilateral Economic Dispatch (MBED) formulation is introduced, allowing for multi-bilateral trading with product differentiation, for instance based on consumer preferences. A Relaxed Consensus+Innovation (RCI) approach is described to solve the MBED in fully decentralized manner. A set of realistic case studies and their analysis allow us showing that such peer-to-peer market structures can effectively yield market outcomes that are different from centralized market structures and optimal in terms of respecting consumers preferences while maximizing social welfare. Additionally, the RCI solving approach allows for a fully decentralized market clearing which converges with a negligible optimality gap, with a limited amount of information being shared.
\end{abstract}


%

\section{Introduction}
\label{sec:intro}

\IEEEPARstart{G}{rowing} climate and environmental concerns have led to a rapid increase of the contribution of renewable energy capacities in electricity generation. These non-dispatchable generators, most often decentralized, put stress on grid operation. In parallel technological advances in data acquisition (smart meters), communication and management have rendered smart housing technologies and internet of things (IoT) possible. Combined with increased storage capabilities (electric vehicles, residential batteries, heat storage, etc.), they allow consumers to be more flexible.  Overall the vertical structure of the system is evolving towards a flatter structure where flexibility and controllability ought to partially shift from generation to consumption. However, alternative approaches and related business models necessary to fully engage the consumption side are not clear yet. Common solutions to harness consumer flexibility, such as aggregators, microgrid management or virtual power plants \cite{nostra2017, burger2017}, have in common that the consumers keep a passive role within the power system. A relevant alternative would be to give consumers, prosumers and more generally all actors of the power system a more proactive role that would eventually turn more beneficial for all. However, this requires profoundly rethinking electricity market design in a more consumer-centric manner, as for instance based on the recently proposed concept of federated power plants \cite{mortsyn18}. An extensive review of the state of the art for consumer-centric electricity markets is available in \cite{sousa2018}.

Different structures for consumer-centric markets are discussed in \cite{parag2016}, from pool-based structures implemented at the micro-grid level \cite{hug2015} to full peer-to-peer approaches. These structures also differ in the degree of centralization in the implementation, depending on whether or not there is a need for a supervisory agent \cite{hug2015, moret17}. Further than going towards more decentralized approaches, made feasible through distributed and consensus-based optimization, it is of utmost importance to propose a market framework allowing all agents to express their preferences. While electricity is commonly seen as an homogeneous good, most often priced under uniform pricing in  forward markets, expressing preferences should yield product differentiation \cite{woo2014} (also referred to as  multi-attribute trading \cite{lai2009}) and hence price differentiation. It is believed that proposing and deploying such novel market structures is a cornerstone for behavioural change of electricity consumers \cite{heiskanen2010}. Consequently, our main contribution here is to describe a peer-to-peer market structure, with a completely decentralized implementation, allowing for product and price differentiation. The underlying dispatch model is coined Multi-Bilateral Economic Dispatch (MBED). A salient feature of the MBED formulation is that it relies on reciprocity constraints for each and every trade to be performed among all agents. Each of these reciprocity constraints may yield a different price for the corresponding electricity exchange.
 
Product (and related price) differentiation is a general concept that allows to place a dynamic value on other aspects of electricity than energy content only. It can be used for instance for an implementation of grid tariffs, for a dynamic tax system, for consumers to express their preferences for some types of generation, and to reflect social aspects. In practice, product differentiation is already possible through bilateral contracts or Power Purchase Agreements (PPA) for big consumers and producers, as well as certificates of origins. However, retailers begin to offer product differentiation to household consumers by gathering differentiated needs (mostly for renewable generation) and negotiating bilateral contracts. The approach of the MBED differs from those in the sense that product differentiation is built in the negotiation mechanism itself. This thus makes that all agents of the power system may account for preferences, different valuations of electricity, differentiated network charges, etc. through the negotiation process leading to the dispatch and pricing of electricity.

To truly obtain a decentralized peer-to-peer market, involving direct interaction and negotiation among all agents of the system, it is proposed to solve the MBED model using consensus-based optimization. Such a decentralized implementation offers benefits such as transparency, data privacy, access to local and updated data and the absence of a central supervisory node. All of those features are expected to help increasing consumer involvement. Our distributed optimization solution approach to the MBED is referred to as the Relaxed Consensus+Innovation method (RCI) since building on the Consensus + Innovation method \cite{kar2012,kar2014}. It is however adapted to multi-bilateral trading by enforcing boundary constraints through Lagrangian relaxation \cite{conejo2006}.

The remaining part of this paper is structured as follows. Section II introduces the general framework for peer-to-peer electricity markets and product differentiation, yielding a general formulation of the MBED problem. Section III describes the RCI method for solving the MBED in a decentralized manner. The workings and benefits of our proposal for a peer-to-peer market combined with a solution approach are illustrated and analyzed in Section IV based a test case of limited size, though realistic. Conclusions and perspectives for future work are gathered in Section V.

\section{Problem Formulation}
\label{sec:formulation}

Let us consider a system with a set $\Omega$ of $N$ agents that are defined as either producers or consumers. The market structure proposed in the following is for a forward market, i.e., to be seen as a resource allocation problem to dispatch and price supply and demand of electric energy. As such, and in line with forward mechanisms, this proposal does not account for (neither accommodates) power system reliability considerations. As for any forward market, it ought to be complemented by mechanisms for reserving and activating ancillary services, as well as by a real-time balancing mechanism which would act as a penalizing instrument for those not being truthful at the forward stage.

\subsection{Peer-to-peer Trading}
\label{sub:variables}
\begin{figure}[ht]
    \centering
    \includegraphics[width=0.95\columnwidth]{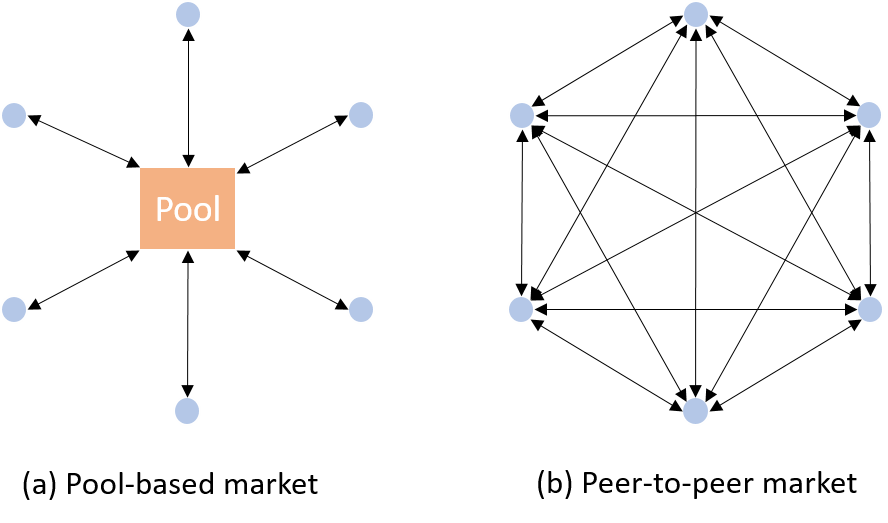}
    \caption{Pool-based (a) and peer-to-peer (b) market structures}
    \label{fig:graph}
\end{figure}

A peer-to-peer architecture for electricity markets is best defined when comparing it to existing pool-based markets (see Figure \ref{fig:graph}, representing the market communication architecture). A peer-to-peer market, characterized by the lack of a supervisory agent, consists of a simultaneous negotiation over the price and energy of multi-bilateral trades along a predefined trading scheme. As such, peer-to-peer structures offer a transparent clearing mechanism that involves prosumers directly while respecting data privacy and being resistant to the failure of any agent. The trading scheme and its underlying graph have a big impact on the complexity and efficiency of the market. Results from graph theory could thus be applied here. For simplicity, only full peer-to-peer markets are considered here, i.e., with a complete graph as in Figure \ref{fig:graph}(b). This graph is for the communication among agents since, in a peer-to-peer market, it is assumed that in principle any agent should be able to negotiate with any other agent. These are market constructs only, regardless of the actual power network (being, e.g., transmission and distribution grids, or possibly a microgrid), similarly to pool-based structures for forward markets, for which actual network constraints and considerations are considered at the balancing stage only. Therefore, when using the term ``neighbor" in the following, this does not necessarily relate to physical location, but to a direct connection on the market communication graph. 

To model this trading scheme, the net power injection $P_n$  of each agent $n\in\Omega$ is split into a sum of bilaterally traded quantities with a set of neighboring agents $m\in\omega_n$, i.e,
\begin{equation}
    P_n=\sumP
\end{equation}
A positive value corresponds to a sale/production and a negative value to a purchase/consumption. The set $\{P_{nm}\mid n\in\Omega,m\in\omega_n\}$ is the set of decision variables. To lighten notations $P_n$ is also used for the whole set of transactions of agent $n$.

The power set-points of an agent $n$ are constrained by the power boundaries $\underline{P_n}$ and $\overline{P_n}$,  
\begin{equation}
    \underline{P_n}\leq P_n\leq\overline{P_n}
\end{equation}
The role of each agent is restrained to either producer or consumer ($\overline{P_n}\underline{P_n}\geq0$) as it simplifies the following derivations. Hence, the decision variables are constrained in sign depending on whether the agent $n$ is a producer ($P_{nm}\geq0$) or a consumer ($P_{nm}\leq0$). However, the approach may be extended to the more general case of prosumers by splitting their buyer (-) and seller (+) role each acting as distinct negotiating entities (although bound by internal constraints) through
\begin{equation}
    P_{nm}=P_{nm}^++P_{nm}^-
\end{equation}


Often, consideration of decentralized resource allocation mechanisms such as our peer-to-peer proposal for electric energy exchange, triggers concerns regarding trust between agents. Such trust concerns maybe related to truthfulness, strategic and malicious behaviour, plus possibly hacking of the negotiation mechanism. Truthfulness may be handled through mechanism design, with the support of a real-time balancing mechanism yielding penalties for those not being truthful at the forward stage. Strategic and malicious behaviour would be similar to the case of current centralized mechanisms, where some agents may attempt to game the market. Similarly for hacking, being centralized or peer-to-peer, such algorithms and software for negotiation and resource allocation may require a layer of protection from cyber-attacks and the likes.

\subsection{Product Differentiation}
\label{sub:Cost_formulation}

The cost function of each agent $n$ is composed of a production cost (or willingness to pay) and a bilateral trading cost. In order to simplify the formulation and understanding of the RCI process, we model the production cost and consumer utility functions as quadratic functions of the power set-point, using three positive parameters $a_n$, $b_n$ and $d_n$, 
\begin{equation}
\label{cost_func}
    C_n(P_n)=\frac{1}{2}a_nP_n^2+b_nP_n+d_n, \quad a_n,b_n,d_n \geq 0
\end{equation}
The reader is referred to \cite{hug2015} for more insights into the geometry of the cost function (\ref{cost_func}). It should be noted that it is fairly common to model such cost and utility functions in a quadratic form, which are seen as realistic for a large class of conventional generators, and certainly some of the best assumption to make when having limited insight in actual utility functions of small consumers and prosumers. However, as long as these functions are convex and with a bijective gradient for the MBED model and the RCI process to be well defined, the overall approach and algorithms described in the present paper are readily applicable. The determination of cost functions, and more generally offering strategies, for all those agents engaging in multi-bilateral trading, is similar to the case of centralized pool structures, for which substantial literature already exists. We expect that the proposal of offering strategies and analysis of strategic behaviour in peer-to-peer electricity markets will be an active topic of investigation in the near future.


The bilateral trading cost is calculated as a linear function of the quantity traded with each neighboring agent,
\begin{equation}
    \Tilde{C_n}(\boldsymbol{p}_n)=\TradingCost
\end{equation}
where $\boldsymbol{p}_n$ is the vector of decision variables of agent $n$ $\boldsymbol{p}_{n} =(P_{nm})_{m\in\omega_n}$ and $c_{nm}$ is the bilateral trading coefficient imposed by agent $n$ on his trade with agent $m$.

The total cost $C$ of the system can thus be expressed as 
\begin{equation}
\label{eq:cost}
    C=\sum_{n\in\Omega}C_n(P_n)+\Tilde{C_n}(\boldsymbol{p}_n)
\end{equation}

The bilateral trading coefficient $c_{nm}$ is expressed in this paper for the purpose of product differentiation and consumer involvement. All trades can be described with a set $\mathcal{G}$ of criteria. The objective valuation of a trade between agents $n$ and $m$ from the perspective of agent $n$ under the prism of criterion $g$ is expressed through a positive parameter $\gamma_{nm}^g$ (called the \textit{trade characteristic} under criterion $g\in\mathcal{G}$). Depending on the criteria, this parameter could be emissions, distance, a rating based on popular vote or on services provided to the community, etc. The  relative costs born by agent $n\in\Omega$ of criterion $g$ is expressed through a parameter $c_n^g$ called the \textit{criterion value} under criterion $g$. Overall the bilateral trading coefficient can be expressed through
\begin{equation}
    c_{nm}=\cnm
\end{equation}
The framework of product differentiation is general and the meaning given to the bilateral trading costs depends in their interpretation. For instance product differentiation can be pushed centrally for a dynamic and specific tax payment or it can be used to better describe consumers' utility through the expression of their preferences. The former implementation may account for taxes, regulatory incentives and network charges, that will then influence the behaviour of market participants and eventual outcomes of the market. The optimal design of these incentives and network charges then becomes a research problem by itself, e.g., to minimize congestion or to support an optimal usage of renewable energy sources. The latter implementation relies on the fact that there is a willingness to pay for certain characteristics of trades for instance for renewable sources of electricity \cite{Guo2014,Strei2015}. In this case, the choice of the bilateral trading coefficient will be part of the broader problem of expressing consumers' utility. However the strategic aspects of this choice for individual consumers have been investigated in \cite{sorin17}, and were shown not to cause welfare distortion over time.

Note that the sign of the criterion value will have an impact on the role of the product differentiation as a rewarding factor or a penalizing factor. Furthermore, to be consistent with a linear implementation of product differentiation, the criterion value should be of opposite signs for sellers and buyers. For instance when modeling prosumers, opposite criterion values should be applied to the split buyer and seller role.

\subsection{MBED Formulation}
\label{sub:formulation}
The equilibrium between production and consumption is represented by a unique balance constraint in a classic pool-based model. It is here replaced by a set of {\it reciprocity constraints} defined for all agents $n\in\Omega$ and $m\in\omega_n$,
\begin{equation}
    P_{nm}+P_{mn}=0
\end{equation}
The Multi-Bilateral Economic Dispatch optimization problem has for objective to maximize the social welfare of the community of agents \eqref{eq:objective} under the constraints of each agent's power injection limits \eqref{eq:boundaries} and sign constraints \eqref{eq:positive}-\eqref{eq:negative} as well as the reciprocity constraints \eqref{eq:balance}. This reads
\begin{subequations}\label{eq:MBED}
\begin{align}
\label{eq:objective}
\min_{D}\quad&\sum_{n\in\Omega}\left(C_n(P_n)+\Tilde{C_n}(\boldsymbol{p_n})\right)&&\\
\label{eq:boundaries}
\text{s.t.} \quad&\underline{P_n} \leq P_n \leq  \overline{P_n} && \forall n \in \Omega\\
\label{eq:balance}
&P_{nm}+P_{mn}=0&&\forall (n,m) \in (\Omega,\omega_n) \\
\label{eq:positive}
&P_{nm}\geq 0 &&\forall (n,m) \in (\Omega_p,\omega_n) \\
\label{eq:negative}
&P_{nm}\leq 0 &&\forall (n,m) \in (\Omega_c,\omega_n)
\end{align}
\end{subequations}
where $D=(\boldsymbol{p}_n\in\mathbb{R}^{\left|\omega_n\right|})_{n\in\Omega}$. $\Omega_p$ and $\Omega_c$ are the sets of producers and consumers, respectively.

The shadow prices of the reciprocity constraints \eqref{eq:balance}, denoted $\lambda_{nm}$, define the prices for the various trades. This implies that all trades can be made at different prices. Differences in price are mostly induced by product differentiation, but can also result from an incomplete trading graph through different price zones. Since the MBED is a convex optimization problem, it only admits a single optimum, which can be obtained by applying a wealth of centralized and distributed optimization methods. 

The simple structure of the MBED described in the above is in line with the type of electricity pools considered for forward markets in Europe, where network constraints are not readily considered (except for cross-zonal transmission limitations), as well as other power system operation aspects related to e.g. voltage levels, reactive power, etc. However, this dispatch may readily be generalized to the network constrained case, while potentially adding security constraints and other power system operation aspects. Considering a DC network-constrained optimal power flow for that dispatch, the nature and properties of the resulting optimization problem will stay the same. Thus, this allows to find an optimal dispatch that is also the true global optimum, while yielding well-defined prices that support the dispatch. Finally, the RCI solution approach described hereafter is readily applicable, since such DC optimal power flow problems are separable and can be solved by a wealth of distributed and decentralized optimization techniques \cite{kargarian18}. Considering an AC network-constrained optimal power flow dispatch would be more difficult though, since modifying the desirable properties of the MBED problem formulation in \eqref{eq:MBED}.

We concentrate here on a single-time-step formulation, in order to cover the basics of our peer-to-peer market structure based on the MBED formulation and product differentiation. However, it may be extended to a multi-time-step formulation to account for, e.g., commitment and ramping constraints and storage characteristics. This may be done by defining and using more complex market products (instead of the single quantity-price products considered here), though at the expense of complexity of the resulting optimization problem and of the properties of the market if the resulting optimization problem becomes non-convex. Alternatively, the MBED could readily account for these temporally binding constraints explicitly, in a way similar to \cite{hug2015}. An exemple generalized formulation of the MBED problem for multiple time steps and temporally-binding constraints is available in \cite{sorin17}.

The MBED model allows for an economic dispatch that respects revenue adequacy, and that allows market participant to be budget balanced. The MBED model is separable along each market participant. Indeed, the agents are bound by the trading reciprocity constraints only, as the quadratic terms of the objective function are independent. Additionally, as Slater's condition is verified, strong duality holds. This opens the way for a decentralized implementation of the MBED using duality. This means that in practice, through the decentralized solution approach described in the following, all involved agents focus only on solving their own local profit maximization (or cost minimization) problem, regardless of overall social welfare. It is by construction, and only by construction, that we know that each agent concentrating on solving their own local problem will yield maximum social welfare solution at convergence. In other words, there is no structural gap between the solutions of the centralized and decentralized approaches to solving the MBED problem, and hence no price to pay for reaching consensus over obtaining the most efficient market outcome.

\section{The Relaxed Consensus + Innovation Solution Approach}
\label{sec:RCI}
The Relaxed Consensus + Innovation (RCI) is a decentralized optimization method inspired by the Consensus + Innovation method presented for instance in \cite{hug2015} in the case of a pool-based market. In the remainder, iterations are indexed with $k$. 

\subsection{Structure of the Method}
The RCI method resembles a dual ascent method as presented for instance in \cite{boyd2011}. The global optimization problem is split into small local problems, the optima of which are reached to obtain the solution of the global problem. Each markets participants aims at solving his own local problem while reacting on the primal and dual estimates of others. The RCI methods is built in such a way that when all local optimum are reached, the equilibrium point is a feasible solution of the global problem which satisfies the first order Karush–Kuhn–Tucker conditions (KKT) and is as such an optimum of the global problem.

The local optimization problem of a given agent $n\in\Omega$ at a given iteration $k$ is
\begin{subequations}\label{eq:local_model}
    \begin{align}
\label{eq:cost_local}
    \min_{D_n}\quad&C_n(P_n)+\Tilde{C_n}(\boldsymbol{p_n})-\boldsymbol{p}_n^\top\boldsymbol{\lambda}_{n}^k&&\\
    \label{eq:local_boundary}
    \text{s.t.} \quad&\underline{P_n} \leq P_n \leq  \overline{P_n} &&\\
    &P_{nm}\geq 0 &&\forall m\in \omega_n\quad \text{if}\quad n\in\Omega_p \\
    \label{eq:local_sign}
    &P_{nm}\leq 0 &&\forall m\in \omega_n\quad \text{if}\quad n\in\Omega_c
    \end{align}
\end{subequations}
where $\boldsymbol{\lambda}_{n}^k =(\lambda_{nm}^k)_{m\in\omega_n}$ is the vector of price estimates of agent $n$ at iteration $k$, $\boldsymbol{p}_n^\top$ is the transposed vector of $\boldsymbol{p}_n$ and $D_n=(\boldsymbol{p}_n\in\mathbb{R}^{\left|\omega_n\right|})$. 
In contrast with to the dual ascent method, the local optimization problem is not solved directly as this may affect the solving of the MBED problem. First of all, solving of the local problem can be computationally costly due to the high number of decision variables per agent. A solution can be to replace that complete solving step with a gradient step and a projection on the feasibility set \cite{nedic2010, hug2015}. Additionally, a direct approach to solve local sub-problems in the MBED will most often results in binary outcomes: when offered different prices, an agent will always try to trade as much as possible from the neighboring agent with the most interesting price and nothing from the others. This results in an undampened oscillating system. 

Our approach is to apply a gradient step. The sign constraints are enforced at each iteration through a closest point projection. Besides, the power boundary constraints are enforced through a Lagrangian relaxation as proposed in \cite{boyd2011} instead of the closest point projection of the C+I method. Using the latter may yield optimality gaps if used for the MBED problem. The corresponding dual variables $\underline{\mu_n}$ and $\overline{\mu_n}$ are estimated using complementary slackness.

\begin{figure}
\begin{tikzpicture}[-latex]

  \matrix (chart)
    [
      matrix of nodes,
      column sep      = 3em,
      row sep         = 5ex,
      column 1/.style = {nodes={treenode}},
      column 2/.style = {nodes={treenode}}
    ]
    {
       Initialize $\boldsymbol{\lambda}_{n}^0$, $\boldsymbol{p}_{n}^0$  &                \\
      {$\lambda$-update\\ $\boldsymbol{\lambda}_{n}^{k+1}$: (\ref{eq:l-update})}   &  \\
      {$\mu$-update\\ $\overline{\mu_n}^{k+1}$: (\ref{eq:mu_up-update}) $\underline{\mu_n}^{k+1}$: (\ref{eq:mu_down-update})}                  &          \\
      {P-update\\ $\boldsymbol{p}_{n}^{k+1}$: (\ref{eq:P-update})} & {Exchange of information \\$F_{nm}^{k+1}$: (\ref{eq:communication_scheme})}\\
      {Stopping criteria\\  \eqref{eq:eps_L}-\eqref{eq:eps_mu}} &       \\
      Market clearing             &       \\
    };
  \draw
    (chart-1-1) edge (chart-2-1)
    \foreach \x/\y in {2/3, 3/4, 4/5} {
      (chart-\x-1) edge (chart-\y-1) }
   (chart-5-1) \yes (chart-6-1)
    (chart-4-2) |- (chart-2-1)
    node[near start, right]{$k:=k+1$};
  \draw[->]
  (chart-5-1) -| (chart-4-2)
      node[near start, below]{no};
\end{tikzpicture}
\caption{Flow chart of the Relaxed Consensus + Innovation algorithm}
\label{fig:flowchart}
\end{figure}
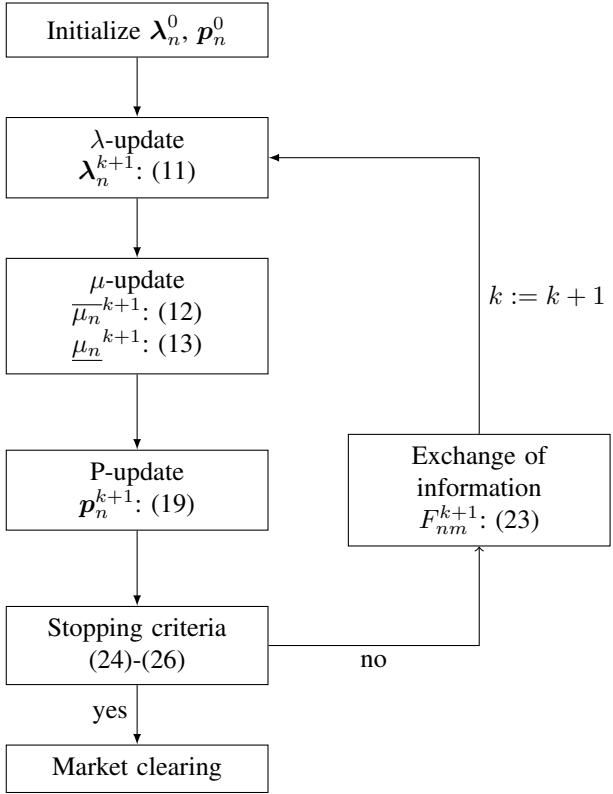

An iteration of the Relaxed Consensus + Innovation method can be split into three steps. The first two steps consist in updating the dual variables of the trading reciprocity constraints and of the power boundary constraints (\ref{sub:dual_update}). The third step consists in updating the decision variables based on a gradient step built on optimality conditions of the local optimality problem (\ref{sub:primal_updates}). Some operational aspects are presented in \ref{sub:Operational}. The structure of the algorithm is summed up in Figure \ref{fig:flowchart}.

\subsection{Dual Updates}
\label{sub:dual_update}
Compared to the dual decomposition of \cite{boyd2011}, the algorithm is here implemented in a fully decentralized setup. The consequence is that price estimates for a given trade are also calculated individually by each participant, even though after convergence, a consensus has to be reached on these estimates (i.e., $\lambda_{nm}=\lambda_{mn}$). This convergence of price estimates is ensured in the price update, (\ref{eq:l-update}), through a \textit{consensus} term. The last term, the \textit{innovation} term, ensures the enforcement of the equality constraint (as in the dual ascent method). This is referred to as the \textit{$\lambda$-update},
\begin{equation}
    \label{eq:l-update}
    \lambda^{k+1}_{nm}=\lambda^{k}_{nm}-\beta^k\left(\lambda^{k}_{nm}-\lambda^{k}_{mn}\right)-\alpha^k\left(P^{k}_{nm}+P^{k}_{mn}\right)
 \end{equation}
where $\alpha^k$ and $\beta^k$ are sequences of positive factors such that any excitation is persistent, so that the series of each sequence diverges \cite{kar2014}. The tuning of these parameters is key to the performance of the algorithm and will usually be a trade-off between convergence speed and resilience to change of setup. The tuning relies as much on the value of the parameters as on the ratio between the two. The performance could be improved by using adaptive factors that adjust this ratio.
 
 The dual variable of the boundary constraints are updated similarly taking into account complementary slackness. This is called the \textit{$\mu$-update},
 \begin{flalign}
    \label{eq:mu_up-update}
    \overline{\mu_n}^{k+1}&=\max(0,\overline{\mu_n}^k+\eta^k(P_n-\overline{P_n}))\\
    \label{eq:mu_down-update}
    \underline{\mu_n}^{k+1}&=\max(0,\underline{\mu_n}^k+\eta^k(\underline{P_n}-P_n))
 \end{flalign}
 where $\eta^k$ is a persistent sequence of positive tuning factors.
 
\subsection{Primal Updates}
\label{sub:primal_updates}
The updates of the decision variables of agent $n$ are based on the KKT optimality conditions of the local optimization problem. The relaxed Lagrangian function of the local optimization problem at iteration $k$ presented in \eqref{eq:local_model} can be expressed as
\begin{equation}
\label{eq:Ln}
\begin{aligned}
L_n^{loc}(\boldsymbol{p}_n,\boldsymbol{\lambda}_n^k,\underline{\mu_n},\overline{\mu_n})&=C_n(\boldsymbol{p}_n)+\Tilde{C_n}(\boldsymbol{p}_n)-\boldsymbol{p}_n^\top\boldsymbol{\lambda}_{n}^k\\
&+\overline{\mu_n}(P_n-\overline{P_n})-\underline{\mu_n}(P_n-\underline{P_n})
\end{aligned}
\end{equation}
With this definition, the first order optimality conditions of the relaxed problem, given by the KKT conditions, are for all agents $n\in\Omega$ and $m\in\omega_n$
\begin{equation}
a_nP_n+b_n+c_{nm}-\lambda_{nm}+\overline{\mu_n}-\underline{\mu_n}=0
\end{equation}
For any trade between two agents $n$ and $m$, we define a new price, referred to as the \textit{perceived price}, through
 \begin{equation}
 \label{eq:perceived}
     \hat\lambda_{nm}=\lambda_{nm}-c_{nm}
 \end{equation}
 By opposition, the trading reciprocity constraint dual variables ($\lambda_{nm}$) are referred to as \textit{trading prices}. As an example, in a tax implementation of product differentiation, the perceived price corresponds to the price after tax, while trading prices correspond to the price before tax. The optimality conditions of the relaxed problem (defined for every decision variable $P_{nm}$ $n\in\Omega, m\in\omega_n$) becomes 
 \begin{equation}
 \label{eq:optimality_relaxed}
\begin{aligned}
a_nP_n+b_n&-\hat\lambda_{nm}+\overline{\mu_n}-\underline{\mu_n}=0
\end{aligned}
\end{equation}
These conditions are equivalent to the optimality condition of a non-zero trade between any two agents $n\in\Omega$ and $m\in\omega_n$ (i.e., when the sign constraint are not binding). Consequently, the perceived prices of a given agent are uniform on the subset of effective trades (trades with $P_{nm}\ne0$ called \textit{non-zero trades}) to a value that equates the marginal cost of production (utility of consumption).  

The Lagrangian of the relaxed problem has bijective gradient when seen as function of the power setpoints. For each negotiated trade (between agent $n$ and $m$, a target power setpoint can be define using the inverse gradient, i.e. 
\begin{flalign}
\label{eq:consumption_target}
    P_n^{(m),k+1}=\frac{\hat\lambda_{nm}-\overline{\mu_n}+\underline{\mu_n}-b_n}{a_n}
\end{flalign}

The primal variables are then updated following \eqref{eq:P-update}, referred to as \textit{P-update} (here for a producer),
\begin{flalign}
    \label{eq:P-update}
    P_{nm}^{k+1}=\max\left(0,P_{nm}^k+f_{nm}^k\left(P_n^{(m),k+1}-P_n^k\right)\right)
\end{flalign}
where $f_{nm}$ is an asymptotically proportional factor defined as
\begin{equation}
\label{eq:ffactor}
    f_{nm}^k=\frac{\left|P_{nm}\right|+\delta^k}{\sum_{l\in\omega_n}(\left|P_{nl}\right|+\delta^k)}
\end{equation}
with $\delta^k$ a positive and persistent sequence. The \textit{max} operator in \eqref{eq:P-update} is used to enforce the sign constraint of the decision variables and is replaced in the case of a consumer by a \textit{min} operator.

Overall, the primal estimates update verifies an averaged optimality condition (if the sign projection is omitted),
\begin{equation}
    \sum_{m\in\omega_n}P_{nm}^{k+1}=\frac{\hat\Lambda_n^{k+1}-\overline{\mu_n}^{k+1}+\underline{\mu_n}^{k+1}-b_n}{a_n}
\end{equation}
where $\hat\Lambda_n^{k+1}$ is an average of the prices perceived by agent $n$ weighted by the factors $f_{nm}^k$,
\begin{equation}
    \hat\Lambda_n^{k+1}=\sum_{m\in\omega_n}f_{nm}^k\hat\lambda_{nm}^{k+1}
\end{equation}

As the perceived prices are uniform on the subset of effective trades and the factor $f_{nm}^k$ is asymptotically proportional to the traded power, after convergence, the averaged optimality condition is equivalent to the optimality condition of the effective trades.


\subsection{Operational Aspects of the Iterative Process}
\label{sub:Operational}
The RCI is a distributed algorithm meant to be implemented in a decentralized fashion, i.e., where all the updates above are done locally by each agent. Consensus-based algorithms like this one are commonly praised for their ability to respect users' privacy and support data security since very little information is shared and only with those agents engaged in multi-bilateral negotiations. All data and related computation are handled locally by each agent. It is unclear whether approaches like inverse optimization or else could allow revealing some of the hidden data and parameters (e.g., preferences) of the market participants. This may be the focus of further work, more generally considering strategic behaviour of agents engaging in multi-bilateral trading.


Here to perform the RCI updates, a minimum set of information is to be shared. As such, at each iteration of the process, the set $F_{nm}^k$ of information sent by an agent $n\in\Omega$ to a neighboring agent $m\in\omega_n$ at iteration $k$ must be the following:
\begin{equation}
\label{eq:communication_scheme}
    F_{nm}^k=\{P_{nm}^k,\lambda_{nm}^k\}
\end{equation}
Interestingly, all the agent's internal production/consumption parameters ($a_n, b_n, \overline{P_n}, \underline{P_n},...$) as well as the criterion parameters ($c_n^g$, $\gamma_{nm}^g$) do not need to be shared in order to reach optimality. The RCI implementation of MBED creates a fully decentralized setup with limited exchange of information, which is a valuable aspect with regards to data privacy. 

The iterative process is stopped when the convergence of the algorithm is established. For that purpose, we define three positive parameters $\epsilon_\lambda$, $\epsilon_P$ and $\epsilon_\mu$ such that the algorithm will terminate when the following conditions are met
\begin{flalign}
        \left|\lambda_{nm}^{k+1}-\lambda_{nm}^{k}\right|&<\epsilon_{\lambda}\label{eq:eps_L}\\
        \left|P_{nm}^{k+1}-P_{nm}^{k}\right|&<\epsilon_{P}\label{eq:eps_P}\\
        \left|\mu_{n}^{k+1}-\mu_{n}^{k}\right|&<\epsilon_{\mu}\label{eq:eps_mu}
\end{flalign}
Criterion \eqref{eq:eps_mu} is optional but can be used for a more precise monitoring of convergence. Similarly, it is common to use a dual convergence criterion only, with the understanding that primal and dual convergence are linked. Further work should be done to understand if it is also the case here.

The number of decision variables that a given agent $n$ deals with is equal to the cardinal of $\omega_n$. This makes this cardinal critical for the complexity of the model and of the algorithm. A full peer-to-peer communication scheme will give a number of variables of the order $N^2$, which renders this model hard to scale up as it is. The topic of scalability, open for future work, will probably be linked to a sparsification of the communication matrix with example structures proposed in \cite{parag2016} such as a multi-level peer-to-peer (a russian-doll structure) or an hybrid peer-to-peer--pool-based structure. 

\section{Application and Case Studies}
\label{sec:cases}
The RCI solving approach to the MBED is evaluated based on simplified though realistic setup as proof-of-concept. After describing the system setup in Section~\ref{sub:ssetup}, a basic illustration of the operation of the RCI and the impact of product differentiation is presented in Section \ref{sub:General_behavior} based on a single time step, to provide the reader with an intuition for its workings. Then, the analysis of our proposal peer-to-peer market is extended to simulation over a full year period (Sections \ref{sub:Convergence_analysis} and \ref{sub:Impact_product_differentiation}) to further look at convergence properties and impact of product differentiation. 

\subsection{Simulation Setup} \label{sub:ssetup}
The setup is composed of 12 agents: 6 producers including two wind turbines, 2 solar PV and 2 fossil generators and 6 consumers including 4 household consumers and 2 industrial consumers. The sequences of wind and solar PV production are taken from \cite{dow2016} and \cite{rat2015}. The consumption sequences as well as the flexibility capacity are derived from \cite{rat2015}. The non time-varying market participants are taken from \cite{hug2015}. Note that compared to \cite{hug2015}, non-zero lower boundary constraints have been added for the generators. The capacities of the generators and consumers are adapted to be comparable. It was chosen to model the PV and the wind turbines as must-take generators ($\overline{P_n}=\underline{P_n}$). This allow do define for them a virtual cost function ($a_n$ and $b_n$) without affecting the optimal solution. Another option would be to model them as zero marginal cost generators ($a_n=b_n=0$ and $\underline{P_n}=0$), however it does not suit the RCI process (non-bijective gradients). The lack of uniqueness of the solutions of the local problem of these agents usually renders the RCI slower than a must-take entity. The agents' parameter for this setup are summed up in the Appendix.

\begin{figure}[ht]
    \centering
    \includegraphics[width=0.92\columnwidth]{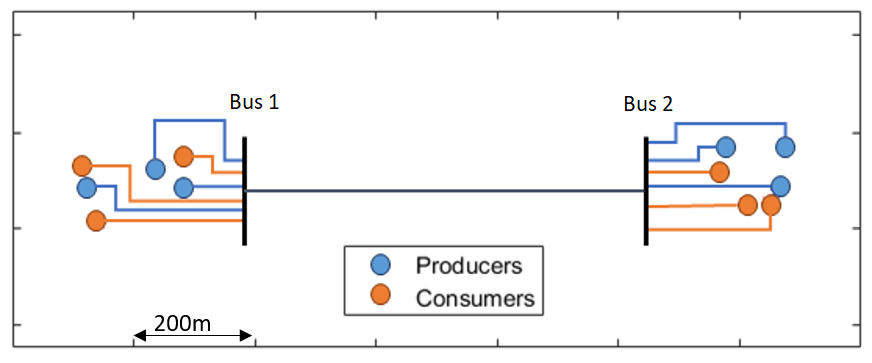}
    \caption{Display of market participants' geographical layout, split into two buses}
    \label{fig:Position}
\end{figure}

The trades are differentiated through a single criterion ($c_{nm}=c_n\gamma_{nm}$) that aims at increasing self and local consumption. For that purpose $\gamma_{nm}$ is chosen to be the euclidean distance between agents $n$ and $m$. The agents are split into two buses (see Figure \ref{fig:Position}), with 3 generators and 3 loads each. Note that the MBED model can be operated without a corresponding network layer. In fact market participants could be connected to completely different part of a nation wide grid. In this test case, however, it was chosen to implement a simplified microgrid structure to clarify the model and to provide quantitative analysis of local consumption.

To take into account the two-bus nature of the setup and the grid structure, the trade characteristics, $\gamma_{nm}$ between any pair of agents of opposite buses is set to a fixed value (1 km in this case) such that every inter-bus trade has the same marginal trading cost. We set the criterion value to a common value in absolute with a negative sign for consumers and positive sign for producers.
The setup is simulated over one year with an hourly time-step. To improve the performance, a warm start based on persistence is used, although it is expected that given the high penetration of renewables and their variability, the persistence will give a high error over the hourly time-step. The construction of the RCI algorithm ensures that, given convergence of the algorithm, the solution found is optimal even though the convergence and its speed depend on the tuning parameters $\alpha^k$, $\beta^k$, $\delta^k$ and $\eta^k$. The parameters chosen after tuning are
\begin{equation}
    \alpha^k=\frac{0.01}{k^{0.01}},\quad\beta^k=\frac{0.1}{k^{0.1}},\quad \eta^k=0.005,\quad\delta^k=1
\end{equation}
The stopping criteria are set to
\begin{equation}
    \epsilon_\lambda=0.001,\quad\quad\epsilon_P=0.01,\quad\quad \epsilon_\mu=0.0001
\end{equation}

\subsection{General Behavior}
\label{sub:General_behavior}
\begin{figure}[ht]
    \centering
    \includegraphics[width=\columnwidth]{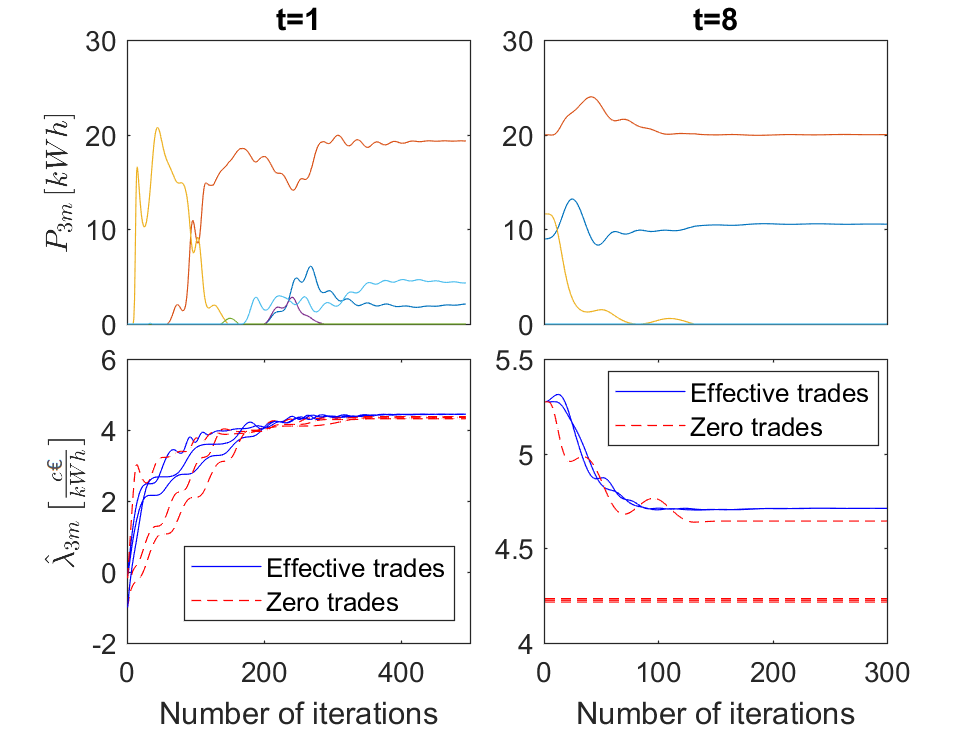}
    \caption{Evolution of energy and perceived prices negotiated by producer 3 at time step 1 and 8.}
    \label{fig:price_power}
\end{figure}

The evolution of the negotiations through the RCI algorithm are plotted for producer 3 in Figure \ref{fig:price_power} to show the convergence behavior. The criterion value is here set to 1 c\euro.kWh$^{-1}$.km$^{-1}$. Two different time steps are shown, one without a warm start (t=1) and one with the warm start based on persistence (t=8). It can be observed that the perceived prices of effective trades are uniform while the other respect the optimality conditions (i.e., the sign constraint dual variables are positive). 

\begin{figure}[ht]
    \centering
    \includegraphics[width=\columnwidth]{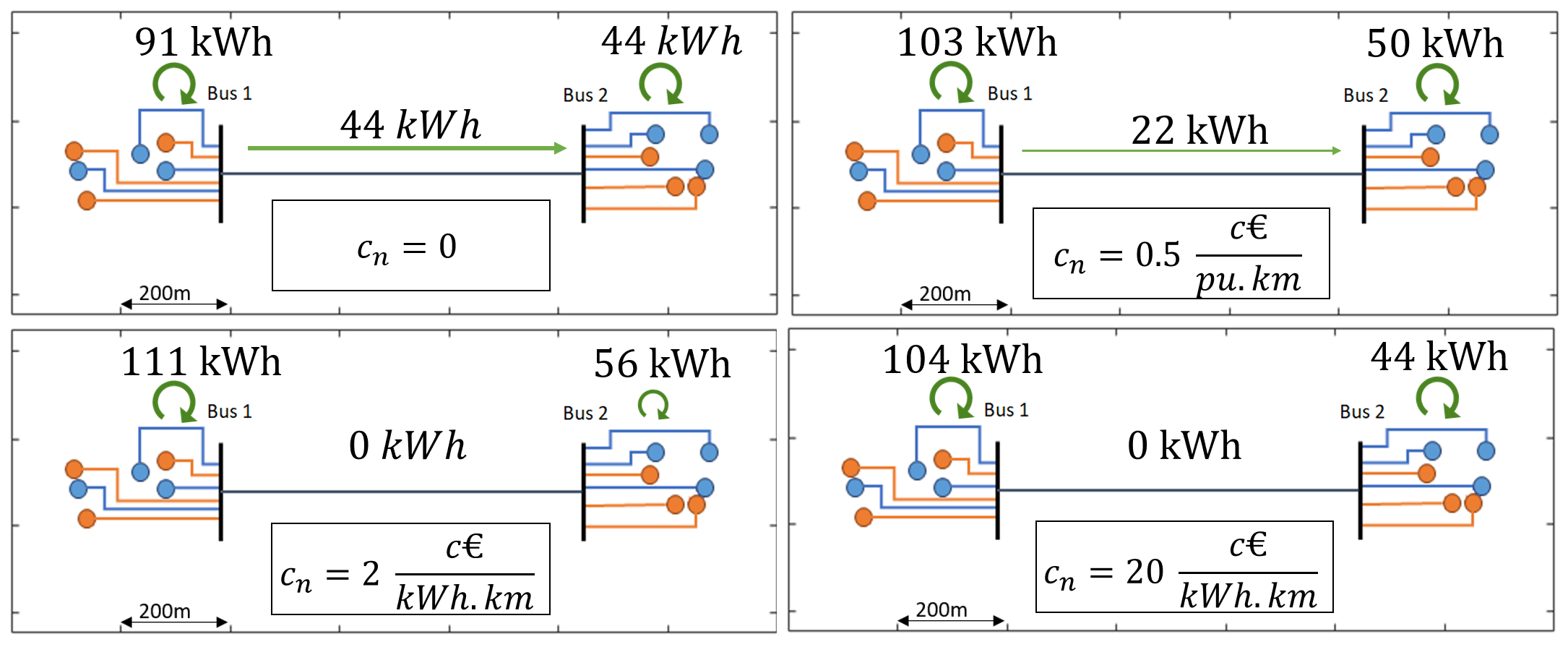}
    \caption{Impact of the common criterion value on the inter-bus exchanges during one time step.}
    \label{fig:Preference}
\end{figure}

A single time-step example is also provided in Figure \ref{fig:Preference} to analyze the impact of product differentiation on grid usage. We take here advantage of the two-bus structure to assimilate grid usage to the use of the inter-bus line.  Without product differentiation ($c_n=0$), the market behaves as a pool-based market with a single power balance constraint and a single price as the trading graph is complete. Consequently, the inter-bus exchanges are high. However the introduction of product differentiation based on distances puts higher marginal bilateral trading costs on inter-bus trades than on intra-bus trades. Consequently, increasing the criterion value has the effect of shifting inter-bus trades toward intra-bus trades until the two buses are autonomous (or all flexibility is used) ($c_n\in\left[0;2\right]$). After that, an increase on the criterion value reduces intra-bus trades, as agents reduce their production/consumption. It can be noted that this impact of product differentiation depends highly on the flexibility capacity of the producers and consumers.

\subsection{Convergence Analysis}
\label{sub:Convergence_analysis}
To certify the optimality of the solution found, a centralized implementation of the MBED was also conducted using a quadratic program solver in Matlab. Both results are compared in terms of objective functions. Note that there is no interest in comparing solutions in terms of decision variable output, as the error on that output will depend on how flat the objective function is around the optimal solution. As the RCI method is based on gradient steps, an optimality gap approach makes more sense.

\begin{figure}[ht]
    \centering
    \includegraphics[width=0.92\columnwidth]{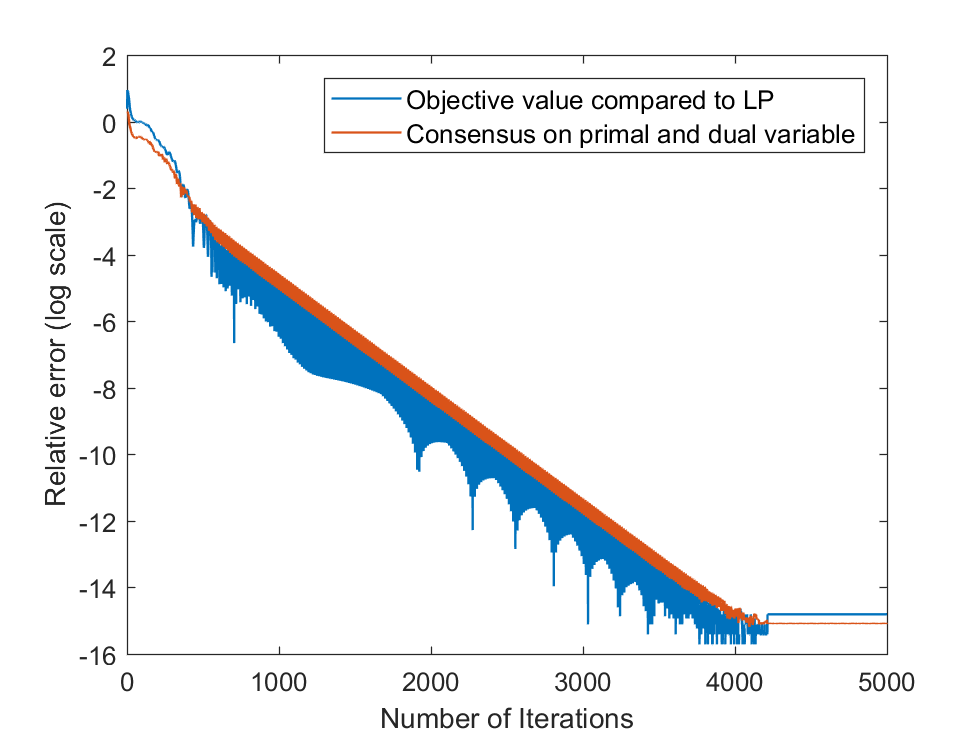}
    \caption{Evolution of the optimality gap and the consensus error throughout the RCI process}
    \label{fig:Optimality}
\end{figure}

The common criterion value is set to 1 c\euro.kWh$^{-1}$.km$^{-1}$. The quality and pace of convergence for a single time-step (t=1) can be seen in Figure \ref{fig:Optimality} through the evolution (without stopping criteria) of optimality gap compared to the optimal solution given by the LP implementation. For an assessment of the feasibility of the solution provided by the RCI, the error on consensus on primal and dual estimates is also displayed. The RCI algorithm presents a logarithmic increase of accuracy of the objective function that is gained in a first phase by approximating the correct solution (here the first 600 iterations) and in a second phase by increasing the accuracy of the solution found. It has to be noted that the performance could be further improved for instance by using adaptive parameters. The convergence speed (mostly for the second phase) can also be significantly improved by increasing the tuning parameters likely at the expense of a more oscillating system which is also less resilient to setup changes. Furthermore an efficient warm start can effectively reduce the first phase. It has been chosen here to use slowly decaying $\alpha$ and $\beta$ parameters, to maintain a high convergence speed and a good resilience, although non-decaying parameters can lead to a faster convergence. 

\begin{figure}[ht]
    \centering
    \includegraphics[width=0.92\columnwidth]{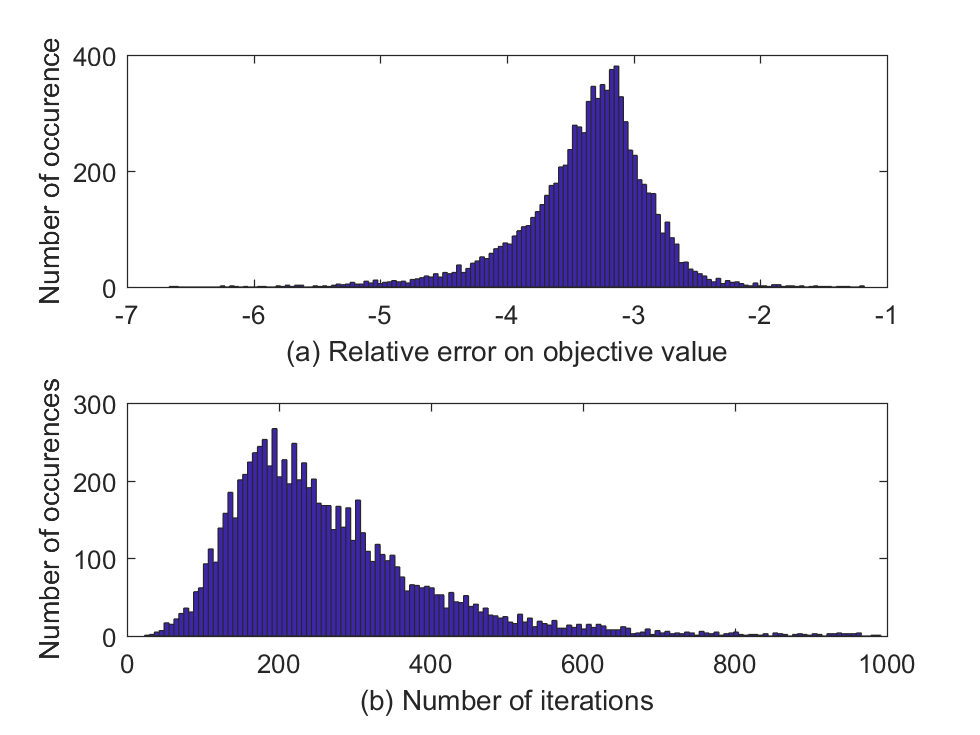}
    \caption{Histogram of the relative error on objective value (a) and the number of iterations to reach convergence (b)}
    \label{fig:histograms}
\end{figure}

The distribution over 1 year of the number of iterations needed to reach the stopping criteria as well as the distribution of the relative error on objective value compared to the LP solution are shown in Figure \ref{fig:histograms} with relative cumulative optimality gap of 0.03$\%$, a maximum relative optimality gap of $4.2\%$ and an average number of iteration to convergence of 298, while the average time to reach the stopping criteria is 0.1s using 64-bit MATLAB R2017a and on an Intel core I7 6500U, 2.5GHz, 8GB RAM.

Employing consensus-based optimization approaches necessarily increases algorithmic complexity and related computational burden if compared to the corresponding centralized solving approach. Scaling to a large number of agents may then become challenging. For an extensive analysis of the properties of the proposed market and related solution approach, as well as a comparison with other proposals for consumer-centric electricity markets, the reader is referred to \cite{moret18b} which looks at 3 types of markets (peer-to-peer, community-based and hybrid) with several hundred agents based on a High-Performance Computing (HPC) implementation.


\subsection{Impact of Product Differentiation}
\label{sub:Impact_product_differentiation}
Two points of view can be used when analyzing the impact of product differentiation in the MBED: the consumer point of view or the system point of view. In the first case, we acknowledge the willingness of market participants to pay for some characteristics of the electricity they trade, and its link to their utility function. We then study this impact in term of social welfare -- with the objective function as defined in \eqref{eq:objective} -- compared to a pool-based market that does not allow for product differentiation. A thorough study of this case would require a deeper analysis of consumer's willingness to pay and it's link to the utility of consuming electricity. In the second case, the product differentiation is introduced as a mean to deal with externalities imposed on power systems (environmental aspects, grid costs...). The differentiation has to be designed to take into account these externalities and we study how product differentiation can effectively reduce them and at which cost.

The second case is implemented here. With product differentiation implemented through a distance criteria, the externality targeted is the grid costs. The topic of this paper is not to ensure that the euclidean distance is a proper description of the marginal impact of a bilateral trade on grid usage nor to describe how grid cost are related to grid usage (in the real-time use or as signal for investments), but rather to show how the use of product differentiation can affect some exogenous parameters that describe externalities or in this case grid usage. Also it is made here the simplifying assumption that grid-usage-related externalities can be effectively described by the energy and the maximal power that transits through the grid. In our two-bus system, the exogenous parameters studied are the energy and maximum power that transits through the line between the two buses as a result of the economic dispatch. 

\begin{figure}[ht]
    \centering
    \includegraphics[width=0.51\textwidth]{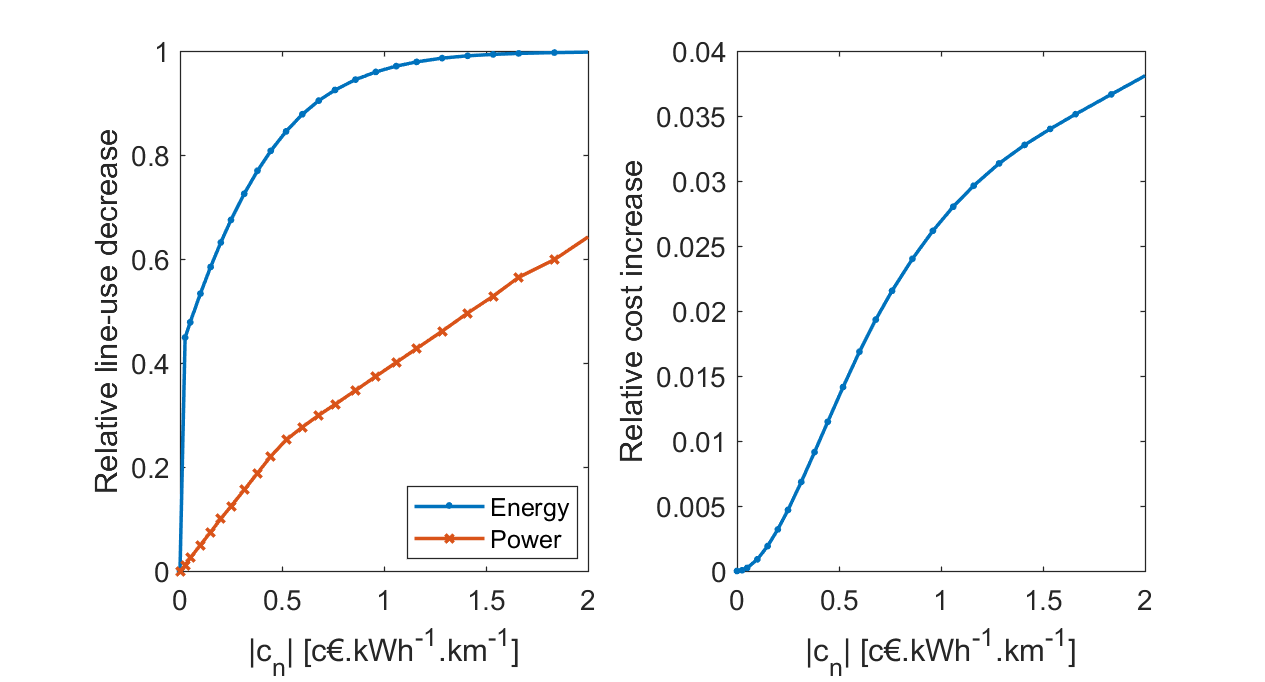}
    \caption{Impact of a common criterion value on the line-use (left) and on costs increase of the MBED (right)}
    \label{fig:Externality_and_cost}
\end{figure}

In Figure \ref{fig:Externality_and_cost} (left), it can be seen how the use of product differentiation based on distances reduces the energy and the maximum power that flows through the inter-bus line over a year compared to a case without product differentiation (equivalent to a pool-based model). A criterion value of 1 c\euro.kWh$^{-1}$.km$^{-1}$ allows to reduce energy flows by more than 95\% and the maximum power by more than 40\%.

\begin{figure}[ht]
    \centering
    \includegraphics[width=0.51\textwidth]{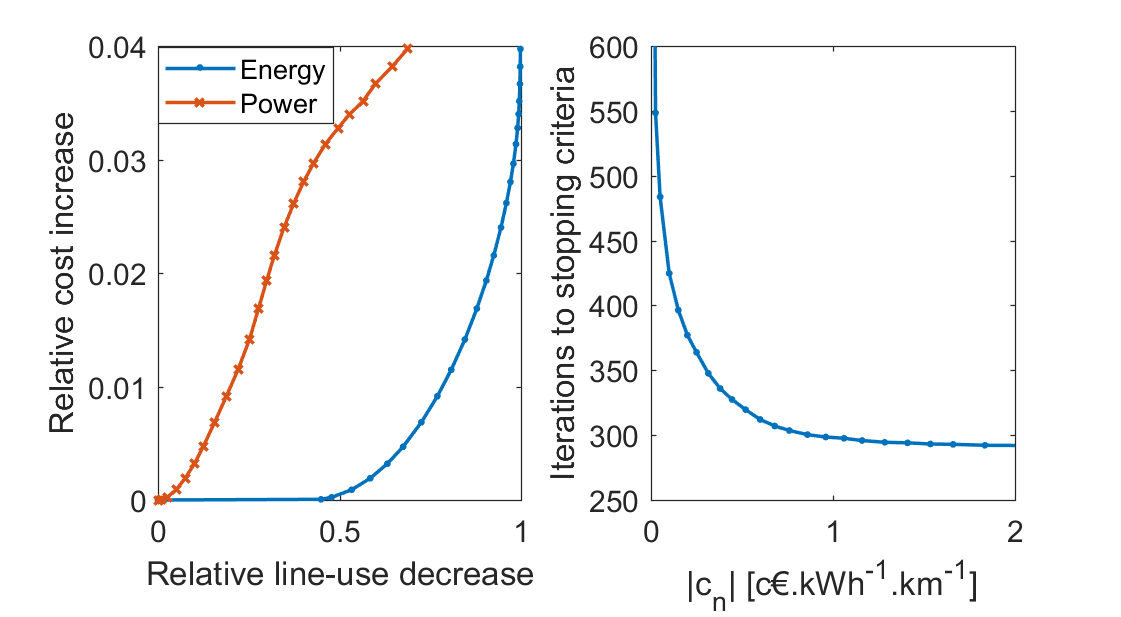}
    \caption{Link between the line-use decrease and the cost increase with the use of product differentiation (left) and the impact of a common criterion value on the iterations to convergence of the RCI (right) }
    \label{fig:iteration_and_combined}
\end{figure}

In the MBED formulation this reduction of externalities is made at the expense of increased cost of production and consumption ($\sum_{n\in\Omega}C_n(P_n)$) called in the rest direct costs. The impact of the criterion value on direct costs is shown in Figure \ref{fig:Externality_and_cost} (right) while the relation between reduction of energy and power flows and cost increase is shown in Figure \ref{fig:iteration_and_combined} (left). it shows that in this setup, almost 50\% of energy flow decrease can be achieved with very low direct cost increase (less than 0.01\%) while 30\% of the power peak  and more than 90\% of energy flows can be reduced with less than 2\% direct cost increase.

From the perspective of the RCI, this reduction in inter-bus trades can also be translated into a reduction of the average number of potential partners which overall can be seen in the decrease of the average number of iteration when the criterion value increases in Figure \ref{fig:iteration_and_combined} (right). Additionally it seems that the iteration number increases very rapidly as the criterion value gets close to zero. This is inherent to the structure of the MBED model: with no product differentiation, the MBED model is equivalent to a pool-based market (given that the communication scheme is complete) whose optimal solution depends only on the total productions/consumptions not on the bilateral trades. It means that in this case the MBED model has multiple solutions which prevents the RCI from finding a solution quickly. The consequence being that, in the case of several market participants not valuing trading costs, the MBED model should be adapted to a hybrid multi-bilateral and pool-based economic dispatch to improve efficiency. The pertinence of these results should be strengthen by a sensitivity analysis for instance on the flexibility capacity of consumers or on the penetration of renewable as well as on the number of buses and market participant.  


\section{Conclusions and Perspectives}
\label{sec:conclusion}

Acknowledging increasing available flexibility of consumers and prosumers, as well as the increasingly decentralized nature of power generation, we have proposed a structure for peer-to-peer electricity markets, based on multi-bilateral trading and product differentiation. We showed that the MBED market framework can be easily and efficiently implemented without the need of a central agent through a distributed Relaxed Consensus+Innovation approach with only a limited exchange of information. This RCI implementation was shown to solve the MBED problem with acceptable optimality gap.

Within our MBED-based framework, product differentiation proved to affect power exchanges in a meaningful way. This framework offers a large variety of implementation. Our proposal decentralized implementation based on consumer preferences is promising in the way that it allows for more pro-active consumer behavior e.g. by favoring local generation or clean generation. Future work regarding product differentiation will focus on the impact of consumer behavior in such a framework. Topic such as heterogeneous criterion values, strategic behavior or impact of free riders will be investigated. The scalability of peer-to-peer markets is generally computationally challenging. One direction for better scalability is the reduction of the number of communication. Its impact of this communication on optimality and convergence speed should be assessed, to eventually study the possibility of communication schemes such as multi-level peer-to-peer markets, as well as hybrid peer-to-peer and pool-based structures.

When it comes to the RCI solution approach to the MBED optimization problem, future work will concentrate on implementing more constrained economic dispatch models. These could either be treated by estimating dual variables through the negotiation process, or by performing a closest-point projection as with the injection bounds in this paper's implementation.
Besides, the MBED problem with product differentiation may allow to readily account for network charges, defined in either endogenous or exogenous manner, which will readily affect multi-bilateral negotiation to account for network topology, see e.g. \cite{baroche18}. Eventually, the MBED approach shall be generalized so as to account for congestion, reactive power compensation, losses, etc. The sharing of these costs in the context of communities is subject to social contracts as investigated in \cite{moret17}. However the interest of P2P schemes will be to readily attribute costs directly induced by the various trades and agents involved, instead of socializing those costs as is most often done in a pool-based market framework.

The peer-to-peer market setup proposed here is for a forward mechanism, hence overlooking some of the salient aspects of today's evolution of power systems and electricity markets, i.e., related to increasing variability and uncertainty due to penetration of renewable generation, changes in demand behaviour, etc. Such issues are not inherent to peer-to-peer electricity markets but instead a general problem to be dealt with in any market setup. In principle, forward markets such as those presented in this paper are not there to account for these aspects. These are dealt with through short-term capacity (i.e., reserve) markets and eventual balancing markets. It is so far unclear whether those other mechanisms should be adapted to the case of peer-to-peer electricity markets. This should be the focus of future work, e.g., concentrating on risk and information sharing, financial products for risk hedging, as well as distributed mechanisms for reserve provision.

\section*{Acknowledgments}

The authors would like to thank Fabio Moret (DTU, Denmark) and Thomas Baroche (ENS Rennes, France) for discussion and feedback on previous versions of the work. The authors would also like to thank four anonymous reviewers for their valuable comments and suggestions which allowed improving the quality of this paper.

%
\appendix 
\subsection{Simulation Setup}
\label{ap:simu}
Tables \ref{tab:generator_parameters_challenging} and \ref{tab:load_parameters_challenging} gather the market participants' utility functions employed. ($P_w^t$) and ($P_s^t$) are the normalized series of respectively wind and solar power generation, while $P_{h}^{\pm,t}$ are the household consumption series with use of upward (+) or downward (-) flexibility, respectively. The wind series are taken from \cite{dow2016} with a installed capacity scalled down to 100 kW per wind turbine. The solar and consumption series are taken from \cite{rat2015} with a installed capacity scalled down to 50 kW per solar installation. The flexibility range is deduced from the controlled generation of \cite{rat2015} while the cost of this flexibility as well as the non-time-varying agents and the virtual cost functions of renewable sources of electricity are deduced from \cite{hug2015}. For a more realistic setup, more effort should be put on a description of flexible generation and its costs settings.

\begin{table}[ht]
\centering
\caption{Parameters of the generators}
\label{tab:generator_parameters_challenging}
\begin{tabular}{|c|c|c|c|c|c|c|}
\hline
\rule{0pt}{3ex}
\multirow{2}{*}{n} & \multirow{2}{*}{Type}    & \multirow{2}{*}{Bus} & $\underline{P_n}$ & $\overline{P_n}$ & $a_n$ & $b_n $ \\[2pt]
&&&[$kWh$]&[$kWh$]&[$\frac{c\text{\texteuro}}{kWh^2}$]&[$\frac{c\text{\texteuro}}{kWh}$] \\[2pt]
\hline
1                 & Wind     & 1       & $P_{w1}^t$     & $P_{w1}^t$    & 0.05  & 3     \\
3                 & Fossil   & 1       & 15                & 105              & 0.056 & 3     \\
6                 & PV & 1       & $P_{s1}^t$      & $P_{s1}^t$     & 0.05  & 3     \\
9                 & Wind     & 2       & $P_{w2}^t$     & $P_{w2}^t$    & 0.05  & 3     \\
10                & Fossil   & 2       & 20                & 90               & 0.06  & 4     \\
12                & PV & 2       & $P_{s2}^t$      & $P_{s2}^t$     & 0.05  & 3     \\ \hline
\end{tabular}
\end{table}
\begin{table}[ht]
\centering
\caption{Parameters of the loads}
\label{tab:load_parameters_challenging}
\begin{tabular}{|c|c|c|c|c|c|c|}
\hline
\rule{0pt}{3ex}
\multirow{2}{*}{n} & \multirow{2}{*}{Type}    & \multirow{2}{*}{Bus} & $\underline{P_n}$ & $\overline{P_n}$ & $a_n$ & $b_n $ \\[2pt]
&&&[$kWh$]&[$kWh$]&[$\frac{c\text{\texteuro}}{kWh^2}$]&[$\frac{c\text{\texteuro}}{kWh}$] \\[2pt]
\hline
2                 & Household     & 1       & -$P_{h1}^{+,t}$  & -$P_{h1}^{-,t}$ & 0.05  & 3     \\
4                 & Household     & 1       & -$P_{h2}^{+,t}$  & -$P_{h2}^{-,t}$ & 0.056 & 3     \\
5                 & Industrial    & 1       &   -120            & -6          & 0.04  & 8     \\
7                 & Household     & 2       & -$P_{h3}^{+,t}$  & -$P_{h3}^{-,t}$ & 0.05  & 3     \\
8                & Household     & 2       & -$P_{h4}^{+,t}$  & -$P_{h4}^{-,t}$ & 0.06  & 4     \\
11                & Industrial    & 2       & -120              & -10             & 0.05 & 8     \\ \hline
\end{tabular}
\end{table}

\subsection{Proof of Strong Duality}

In Section~\ref{sub:formulation}, it is stated that Slater's condition is verified, and from that is deduced that strong duality holds. This appendix aims at clarifying this statement.

The cost functions are quadratic with a positive quadratic coefficient. Similarly the constraints are linear thus convex. This implies that the MBED is a convex optimization process. The feasibility of the MBED problem depends on the feasibility of the corresponding pool-based model where the balance constraints (equations \eqref{eq:balance}) is replaced by a global balance constraint:
\begin{equation}
    \sum_{n\in\Omega}P_n=0
\end{equation}

If the market is well designed (meaning if there is adequate means of production and consumption), the pool-based model is feasible. This implies that the MBED is feasible. Indeed if $x^*=(P_n^*,n\in\Omega)$ is a feasible solution of the pool-based problem, then $y^*=(P_{nm}^*,n\in\Omega,m\in\omega_n)$ is a feasible solution of the MBED problem where $y^*$ is defined such that:
\begin{flalign}
        P_{nm}^*=\frac{P_m^*}{\sum_{l\in\Omega_c}P_l^*}P_n^*&\quad\quad\forall n\in\Omega_p,m\in\Omega_c\\
        P_{nm}^*=-P_{mn}^*&\quad\quad\forall n\in\Omega_c,m\in\Omega_p
\end{flalign}

As a convex and feasible optimization problem, Slater's condition tells us that strong duality holds.



\begin{thebibliography}{}

\bibitem{nostra2017}S. M. Nosratabadi, R.-A Hooshmand, and E. Gholipour, “A comprehensive review on microgrid and virtual power plant concepts employed for distributed energy resources scheduling in power systems," \emph{Renew. Sust. Energ. Rev.}, vol. 67, pp. 341-363, 2017.

\bibitem{burger2017}S. Burger,  J.P. Chaves-Ávila, C. Batlle, and I.J. Pérez-Arriaga, “A review  of  the  value  of  aggregators  in  electricity  systems,” \emph{Renew. Sust. Energ. Rev.}, vol. 77, pp. 395–405, 2017

\bibitem{mortsyn18} T. Mortsyn, N. Farrell, S.J. Darby and M. McCulloch, “Using peer-to-peer energy-trading platforms to incentivize prosumers to form federated power plants,” \emph{Nat. Energy}, vol. 3, pp. 94-101, 2018.

\bibitem{parag2016} Y. Parag and B.K. Sovacool, “Electricity market design for the prosumer era,” \emph{Nat. Energy}, vol. 1, art. no. 16032, 2016.

\bibitem{hug2015} G. Hug, S. Kar, and C. Wu, “Consensus + Innovations approach for distributed multi-agent coordination in a microgrid,” \emph{IEEE Trans. Smart Grid}, vol. 6, no. 4, pp. 1893-1903, 2015.

\bibitem{sousa2018} T. Sousa, T. Soares  P. Pinson, F. Moret, T. Baroche, and E. Sorin, "Peer-to-peer and community-driven markets: a comprehensive review," under review, available at: \url{http://pierrepinson.com/docs/Sousaetal2018.pdf}, 2018. 

\bibitem{moret17} F. Moret and P. Pinson, “Energy Collectives: A community and fairness based approach to future electricity markets,” \emph{IEEE Trans. Power Syst.}, available online, 2018.

\bibitem{woo2014}C.K. Woo, P. Sreedharan, J. Hargreaves, and F. Kahrl, “A review of electricity product differentiation,” \emph{Appl. Energy}, vol. 114, pp. 262-272, 2014.

\bibitem{lai2009}G. Lai and K. Sycara, “A generic framework for automated multi-attribute negotiation,” \emph{Group Decis. Negot.}, vol. 18, no. 2, pp. 169-187, 2009.

\bibitem{heiskanen2010} E. Heiskanen, M. Johnson, S. Robinson, E. Vadovics, and M. Saastamoinen, “Low-carbon communities as a context for individual behavioural change,” \emph{Energ. Policy}, vol. 38, no. 12, pp. 7586–7595, 2010.

\bibitem{kar2012} S. Kar and G. Hug, “Distributed robust economic dispatch in power systems: A Consensus + Innovations approach,” \emph{IEEE Power and Energy Society General Meeting}, San Diego, CA, 2012.

\bibitem{kar2014} S. Kar, G. Hug, J. Mohammadi, and J. M. Moura, “Distributed state estimation and energy management in smart grids: A Consensus ${+} $ Innovations approach”, \emph{IEEE J. Sel. Top. Signal Proc.}, vol. 8, no. 6, pp. 1022–1038, 2014.

\bibitem{conejo2006}A. J. Conejo, F. J. Nogales, and F. J. Prieto, \emph{Decomposition Techniques in Mathematical Programming}, \emph{Springer}, New York, NY, USA: Springer-Verlag, 2006.

\bibitem{Guo2014} X. Guo, H. Liu, X. Mao, J. Jin, D. Chen, and S. Cheng, “Willingness to pay for renewable electricity: A contingent valuation study in Beijing, China,” \emph{Energ. Policy}, vol. 68, pp. 340-347, 2014.

\bibitem{Strei2015} D. Streimikiene and A. Balezentis, “Assessment of willingness to pay for renewables in Lithuanian households,” \emph{Clean Technol. Envir.}, vol. 17, no. 2, pp. 515–531., 2015.

\bibitem{sorin17} E. Sorin, "Peer-to-peer electricity markets with product differentiation -- Large-scale impact of a consumer-oriented market," M.Sc. thesis report. Technical University of Denmark, 2017.

\bibitem{kargarian18} A. Kargarian, J. Mohammadi, J. Guo, S. Chakrabarti, M. Barati, G. Hug, S. Kar, and
R. Baldick, “Toward distributed/decentralized DC optimal power flow implementation in future electric power systems,” \emph{IEEE Trans. Power Syst.}, available online, 2018.

\bibitem{boyd2011} S. Boyd, N. Parikh, E. Chu, B. Peleato, and J. Eckstein, “Distributed optimization and statistical learning via the alternating direction method of multipliers,” \emph{Found. Trends Mach. Learn.}, vol. 3, no. 1, pp. 1–122, 2011.






\bibitem{nedic2010} A. Nedic, A. Ozdaglar, and P.A. Parrilo, “Constrained consensus and optimization in multi-agent networks,” \emph{IEEE Trans. Autom. Control}, vol. 55, no. 4, pp. 922–938, 2010.

\bibitem{dow2016} J. Dowell and P. Pinson, “Very-short-term probabilistic wind power forecasts by sparse vector autoregression,” \emph{IEEE Trans. Smart Grid}, vol. 7, no. 2, pp. 763-770, 2016.

\bibitem{rat2015} E. Ratnam, S. Weller, C Kellett, and A. Murray, “Residential load and rooftop PV generation: an Australian distribution network dataset,” \emph{Int. J. Sust. Energy}, vol. 36, no. 8, pp. 787–806, 2015.

\bibitem{moret18b} F. Moret, T. Baroche, P. Pinson, and E. Sorin, "Negotiation algorithms for peer-to-peer electricity markets: Computational properties," in \emph{Proc. Power Syst. Comput. Conference 2018}, Dublin, Ireland, June 2018.

\bibitem{baroche18} T. Baroche, P. Pinson, R. Le Goff Latimier., and H. Ben Ahmed, "Exogenous approach to grid cost allocation in peer-to-peer electricity markets," arXiv preprint, arXiv:1803.02159.

\end{thebibliography}
\end{document}